\documentclass[lettersize,journal]{IEEEtran}
\usepackage{amsmath,amsfonts}
\usepackage{algorithmic}
\usepackage{algorithm}
\usepackage{array}
 \usepackage{subfigure}
\usepackage{textcomp}
\usepackage{url}
\usepackage{verbatim}
\usepackage{graphicx}
\usepackage{cite}
\begin{document}

\title{ \huge {VHetNets for AI and AI for VHetNets: An Anomaly Detection Case Study for Ubiquitous IoT}}

\author{Weili~Wang,
         Omid~Abbasi,
         Halim~Yanikomeroglu,
         Chengchao~Liang,
        Lun~Tang,
        and Qianbin~Chen

%\author{IEEE Publication Technology,~\IEEEmembership{Staff,~IEEE,}
        % <-this % stops a space
\thanks{W. Wang, C. Liang, L. Tang and Q. Chen (corresponding author) are with the School of Communication and Information Engineering and the Key Laboratory of Mobile
Communication, Chongqing University of Posts and Telecommunications, Chongqing 400065, China.}% <-this % stops a space

\thanks{O. Abbasi and H. Yanikomeroglu are with the Department of Systems and Computer Engineering, Carleton University, Ottawa, ON, Canada.}}



\maketitle

\begin{abstract}
Vertical heterogenous networks (VHetNets) and artificial intelligence (AI) play critical roles in 6G and beyond networks. This article presents an AI-native VHetNets architecture to enable the synergy of VHetNets and AI, thereby supporting varieties of AI services while facilitating automatic and intelligent network management. Anomaly detection in Internet of Things (IoT) is a major AI service required by many fields, including intrusion detection, state monitoring, device-activity analysis, security supervision and so on. Conventional anomaly detection technologies mainly consider the anomaly detection as a standalone service that is independent of any other network management functionalities, which cannot be used directly in ubiquitous IoT due to the resource-constrained end nodes and decentralized data distribution. In this article, we develop an AI-native VHetNets-enabled framework to provide the anomaly detection service for ubiquitous IoT, whose implementation is assisted by intelligent network management functionalities. We first discuss the possibilities of VHetNets used for distributed AI model training to provide anomaly detection service for ubiquitous IoT, i.e., VHetNets for AI. After that, we study the application of AI approaches in helping provide automatic and intelligent network management functionalities for VHetNets, i.e., AI for VHetNets, whose aim is to facilitate the efficient implementation of anomaly detection service. Finally, a case study is presented to demonstrate the efficiency and effectiveness of the proposed AI-native VHetNets-enabled anomaly detection framework.
\end{abstract}

%\begin{IEEEkeywords}
%Article submission, IEEE, IEEEtran, journal, \LaTeX, paper, template, typesetting.
%\end{IEEEkeywords}

\section{Introduction}
6G and beyond networks are expected to exhibit some unique characteristics. First, satellites in low, medium and geostationary Earth orbits, high altitude platform stations (HAPSs), unmanned aerial vehicles (UAVs) and terrestrial base stations are integrated into vertical heterogenous networks (VHetNets) to provide a global coverage \cite{13-9749175}. Second, ubiquitous Internet of Things (IoT), which is playing an important role in varieties of verticals including smart homes, healthcare wearables, environmental monitoring, industrial control, agriculture and transportation, supports seamless connectivity anytime, anywhere, and for everything \cite{2-9606808,5-9535454}. Third, intelligence exists in every corner of networks, ranging from end devices to the central network controller. Numerous network nodes are endowed with built-in artificial intelligence (AI), thereby supporting varieties of AI services while facilitating automatic and intelligent network management \cite{14-9749222}.

Anomaly detection is defined as a process of automatically detecting whether devices, components or systems are in their normal running states or not \cite{6-9301388}. Anomaly detection is one of the indispensable functionalities in 6G ubiquitous IoT system for the following three reasons:
\begin{itemize}
\item 6G networks are envisioned to support new services while satisfying their various and stringent QoS requirements, including ultra-low latency and ultra-high reliability required by autonomous driving and industrial control systems \cite{14-9749222}. An efficient anomaly detection framework is vital to detect or even predict possible abnormal running states happened in these systems.
\item Ubiquitous IoT devices provide many opportunities for malicious attackers to degrade their performance including denial of service, botnet, collusion, and many other types of attacks. Anomaly detection could build a model representing normal running states and identify abnormal behaviors or attacks based on the deviation from the model.
\item Anomaly detection is the first step to recover networks or systems from failures and compensate the degraded service performance in an automatic and intelligent way, which is expected to realize in 6G network management.
\end{itemize}

Sensing and communication are two fundamental functionalities for the implementation of anomaly detection in ubiquitous IoT. Traditional anomaly detection schemes in IoT usually utilize static sensors to perform data sensing and IoT gateways to forward the sensing data to a central server for storage and analysis \cite{3-9543534}. In this case, sensing and communication are separately accomplished by two standalone modules, which causes high hardware cost, power consumption and signaling latency. Besides, in situations where sensing targets are placed in hard-to-reach regions, such as hazardous or poisonous locations, traditional sensing solutions would be invalid. The integrated sensing and communication design has been recognized to be indispensable in ubiquitous IoT due to its low hardware cost, power consumption and signaling latency \cite{18-9606831}. For implementing anomaly detection in ubiquitous IoT, HAPSs and UAVs with integrated sensing and communication capacity  are advanced approaches for smart sensing and data collection \cite{2-9606808,3-9543534}.

Machine learning (ML) is the most commonly used technique to obtain the anomaly detection model. Based on adopted ML approaches, anomaly detection models are classified as follows: clustering-based models (such as K-means, DBSCAN, and local outlier factor), classification-based models (such as K-nearest neighbor, support vector machine, and Bayesian network), dimension-reduction-based models (such as principal component analysis (PCA), and recursive PCA), and hybrid models that combine multiple approaches together \cite{4-9116088}.  Existing works mainly consider the anomaly detection as a standalone service that is independent of any other network management functionalities, such as network planning, device association, resource scheduling, and so on. However, in ubiquitous IoT system, the ML-based anomaly detection cannot be implemented without the allocated computing resources or sensing data from the associated devices. Therefore, we propose an AI-native VHetNets-enabled anomaly detection framework to provide the anomaly detection service for ubiquitous IoT, whose implementation will be assisted by intelligent network management functionalities.

\begin{figure*}[htbp]
\centering
\includegraphics[width=5in]{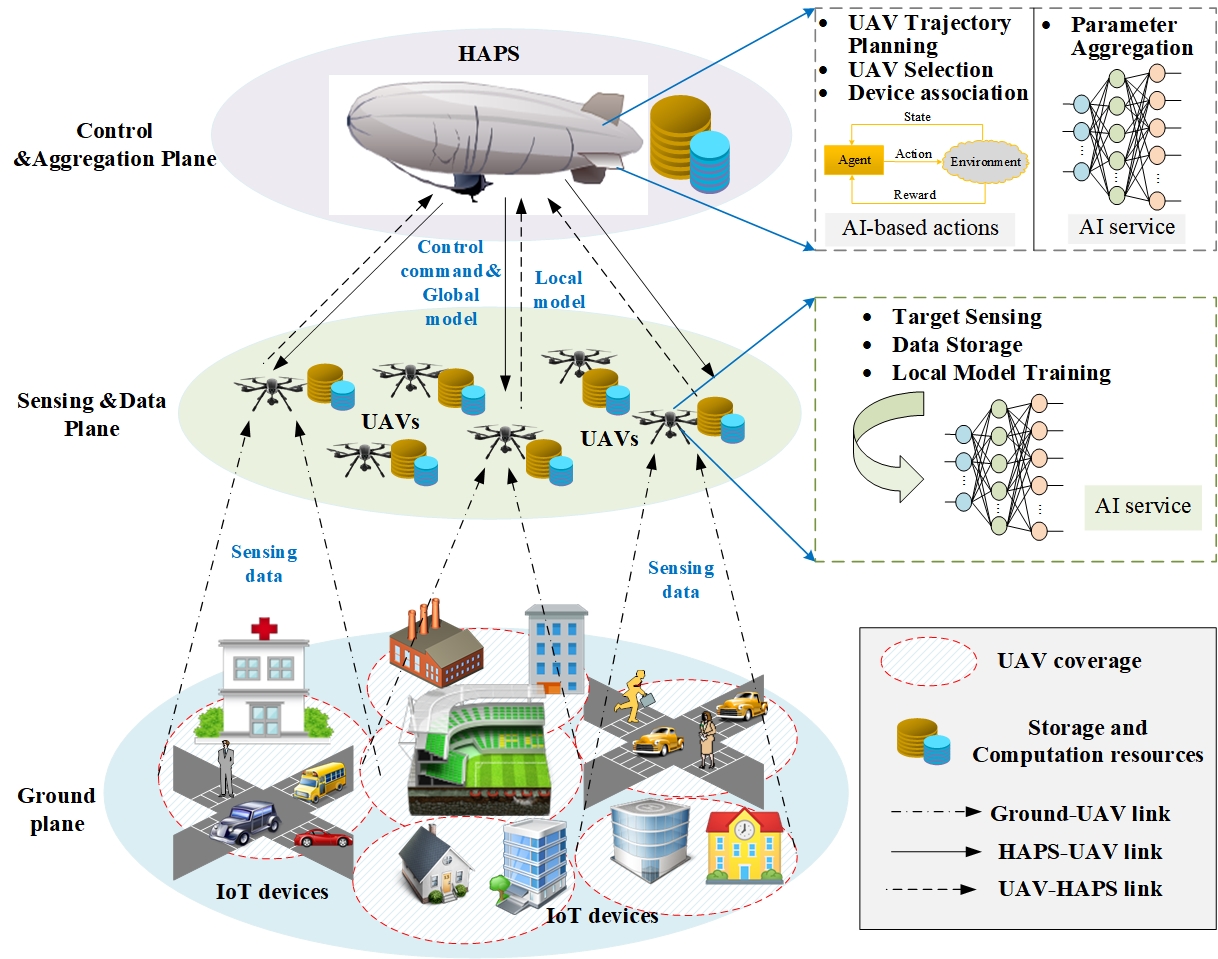}
\caption{VHetNets for AI and AI for VHetNets.}
\end{figure*}

AI-native means that, as a built-in component in VHetNets, AI exists not only in VHetNets as services for the implementation of anomaly detection, but also in the network controller for providing automatic and intelligent network management. Therefore, the synergy of AI and VHetNets is two-fold.  On one hand, HAPS and UAVs in VHetNets can work as agents for distributed AI model training to provide the anomaly detection service, namely VHetNets for AI. On the other hand, AI techniques can be applied to provide automatic and intelligent network management functionalities (such as network planning, device association, and resource scheduling) in VHetNets, namely AI for VHetNets. Moreover, the two components of AI-native VHetNets, namely VHetNets for AI and AI for VHetNets, are not independent but complementary to each other. The former requires the latter to provide adequate computing and bandwidth resources for the implementation of anomaly detection service while the latter relies on results of the former (such as finding degraded devices) to improve the network management efficiency. The contributions of this article are summarized as follows:
\begin{itemize}
\item Develop an AI-native VHetNets-enabled framework to provide the anomaly detection service for ubiquitous IoT, which is implemented through the hierarchical collaboration among three layers, namely HAPS, UAVs and ground IoT devices.
\item Establish a distributed anomaly detection model based on asynchronous federated learning (AFL), where UAVs with integrated sensing and communication capacity are responsible for the data sensing, storage and local model training, and HAPS is responsible for the local model aggregation. Local models are aggregated at HAPS asynchronously considering UAVs' mobility and heterogeneous remaining energy. The AFL-based anomaly detection model utilizes the communication, computing and storage resources in VHetNets to conduct the distributed AI model training.
\item Design a deep reinforcement learning (DRL)-based network management scheme to realize AI for VHetNets through dynamic interactions with the network environment. Because the adopted actions for network management include both continuous and discrete variables, a compound-action actor-critic (CA2C) method is introduced to combine the advantages of the deep deterministic policy gradient (DDPG) for handling continuous action spaces and the deep Q-network (DQN) for handling discrete action spaces.The CA2C-based network management scheme will assist the implementation of the AFL-based anomaly detection model.
\end{itemize}

In this article, we develop an AI-native VHetNets-enabled anomaly detection framework to provide the anomaly detection service for ubiquitous IoT devices. First, the architecture of the AI-native VHetNets-enabled anomaly detection framework is introduced. Second, we present the techniques of VHetNets for AI to implement anomaly detection and AI for VHetNets to provide intelligent and automatic network management functionalities. Then, we validate the efficiency and effectiveness of the proposed framework through a case study. Finally, the article is concluded and the future research is presented.

\section{AI-native VHetNets-enabled anomaly detection framework}
Traditional anomaly detection framework usually depends on a centralized server to collect all raw data and train a global AI model based on the collected data. However, due to the data explosion in ubiquitous IoT, the privacy concern and limited communication resources for data transmission have invalidated the centralized framework \cite{22-9453811}. As an emerging decentralized AI framework, federated learning (FL) is effective to conquer the above challenges by training a shared model in distributed participating agents \cite{2-9606808}.

As shown in Fig. 1, the proposed AI-native VHetNets-enabled anomaly detection framework is established on the FL and leverages the hierarchical collaboration among three layers, namely HAPS, UAVs and ground IoT devices, whose main roles are elaborated as follows:

\textbf{Ground IoT devices.} Ground IoT devices are kinds of applications in a variety of verticals, from industry to environments, from transportation to healthcare, from home area to public venues, and so on. By collecting and analyzing the sensing data indicating running states of different systems, anomaly detection model can be constructed to detect abnormal behaviors of monitored systems and provide early warnings before the occurrence of failure.

\textbf{UAVs.} Each UAV with integrated sensing and communication capacity plays roles of a sensor, computing node and storage node. It is responsible for sensing ground IoT devices within its coverage and training the local anomaly detection model based on its sensing data. Using UAVs as aerial nodes to provide wireless sensing support from the sky is a promising paradigm for its three advantages. First, its elevated altitude and reduced signal blockage contribute to a wider field of view for the UAV-based sensing compared to ground sensors. Second, the highly flexible and controllable 3D UAV mobility facilitates the deployment of UAV sensors in hard-to-reach regions, such as hazardous or poisonous locations. Finally, the high mobility of UAVs allows the performance optimization of sensing through dynamic UAV trajectory planning \cite{20-9456851}.

Traditional FL commonly uses a synchronous learning framework to aggregate the local model parameters from all participating agents. Due to UAVs' mobility and heterogeneous remaining energy, waiting for all UAVs to finish their local training before aggregation will inevitably increase the global learning delay. Therefore, AFL is utilized among UAVs to improve the learning efficiency.

\textbf{HAPS.} A HAPS is a quasi-stationary network node that operates in the stratosphere at an altitude of around 20 km \cite{21-9380673}. HAPS, which provides line-of-sight communication and wide coverage with a radius of 60-500 km, has been regarded as an indispensable component for 6G networks. By adopting the AI technique of deep reinforcement learning, HAPS can control and manage the whole network in an intelligent and automatic way through unintermittent interactions with the network environment. In the proposed AI-native VHetNets-enabled anomaly detection framework, HAPS is responsible for the dynamic UAV trajectory planning, UAV selection for AFL, device association strategies, and the parameter aggregation uploaded by UAVs.

\section{VHetNets for AI: An AFL-based anomaly detection scheme for ubiquitous IoT}
Instead of uploading all sensing data from UAVs to HAPS and training the anomaly detection model at HAPS in a centralized manner, AFL enables selected UAVs to execute local training on their own sensing data and avoid raw data transmission to HAPS for privacy preservation and communication resource savings. Specifically, each UAV uploads its local model to HAPS periodically if it is selected to participate in this global aggregation round. After that, HAPS aggregates local models and broadcast the global model to all associated UAVs for another global training round.

AFL framework needs an efficient UAV selection strategy to decrease the effect of UAVs with low energy and low quality of model to the global model's learning efficiency and accuracy.  The UAV selection strategy could be explored by the DRL technique elaborated in the next section.

Network anomalies are usually rare events, which will cause the historical sensing data forming an imbalanced dataset. Therefore, we implement the anomaly detection through building the normal behavior model and determining abnormal behaviors according to their degree of deviation from normal ones. In this article, we use the generative adversarial network (GAN) to capture the distribution of normal data in UAVs for its powerful ability in capturing the distribution from high-dimensional complex real-world data \cite{30-8253599}.

\textbf{GAN}. GAN consists of two models: the generator and the discriminator, which are usually built with neural networks. The generator attempts to generate new (fake) samples from latent variables to fool the discriminator, and the discriminator attempts to distinguish fake samples from real ones. By training the generator and discriminator through the adversarial learning, the Nash equilibrium can be achieved theoretically. After convergence, the generator can generate fake samples sharing the same distribution with real ones and the discriminator is unable to distinguish them. Fig. 2 illustrates the training processes of the AFL-based anomaly detection scheme assisted by GAN, whose workflow consists of nine steps as follows:

\begin{figure*}[htbp]
\centering
\includegraphics[width=5.5in]{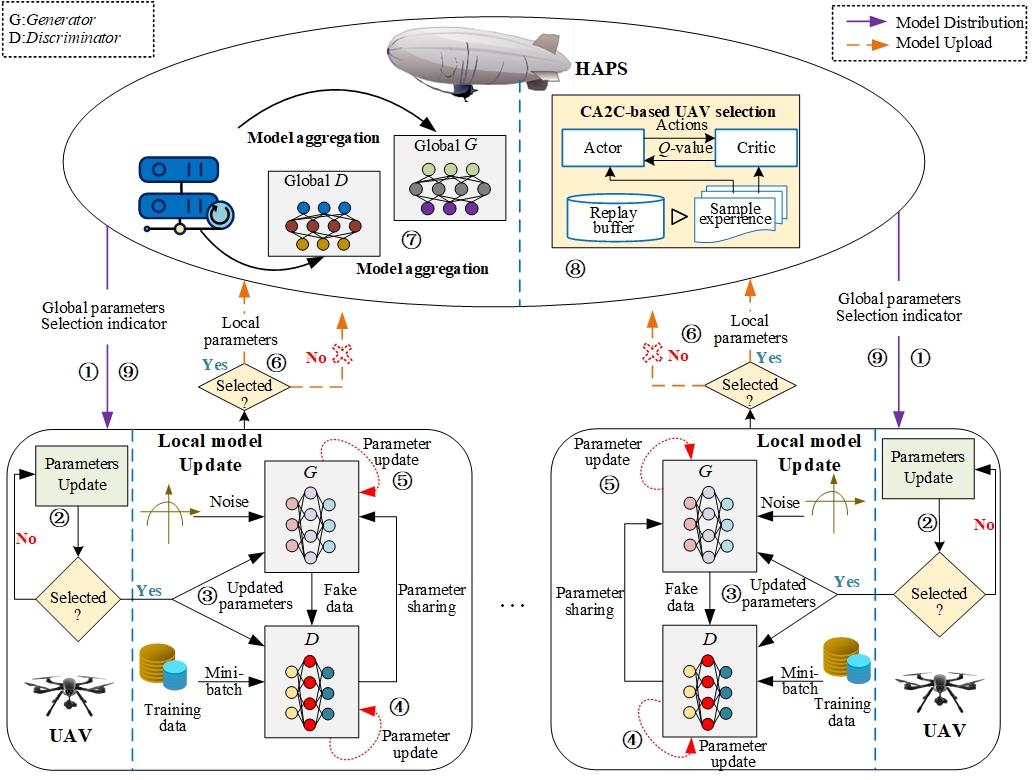}
\caption{Training processes of the AFL-based anomaly detection scheme assisted by GAN.}
\end{figure*}

\begin{enumerate}
    \item HAPS distributes initial global model parameters and UAV selection indicators to each UAV as local model parameters and update indicators, respectively;
    \item Each UAV sets its local model parameters as global ones, and then checks the selection indicator to determine whether to implement the local training step or not;
    \item If a UAV is selected to participate in this aggregation round, then activate training processes of the local generator and discriminator;
    \item Sample a mini-batch of fake data and real data to train the local discriminator;
    \item Based on the local discriminator model, sample a mini-batch of noise to train the local generator every $N$ discriminator training rounds;
    \item Upload local parameters to HAPS for model aggregation every $K$ local training rounds;
    \item HAPS obtains global generator and discriminator models by the federated averaging;
    \item HAPS implements the DRL-based UAV selection to determine the optimal UAV subset for the next global aggregation round;
    \item HAPS distributes model parameters and UAV selection indicators to all UAVs for their further updates.
\end{enumerate}

The above steps are implemented repeatedly until global models converge. After the global generator and discriminator are well trained, an anomaly score, which is defined as the weighted combination of the generator loss and discriminator loss, can be constructed to measure the abnormality of new sensing data.

%This method's workflow consists of nine steps as follows: \textcircled {1} HAPS distribute initial global model parameters and UAV selection indicators to each UAV as the local model parameters and  update indicators, respectively;
%\textcircled {2} Each UAV sets its local model parameters as the global ones, then check the selection indicator to determine whether to implement the local training step or not;
%\textcircled {3} If a UAV is selected to participate in this training round, then activate the training process of local generator and discriminator;
%\textcircled {4} Sample a mini-batch of fake date and real data to train the discriminator;
 %\textcircled {5} Sample a mini-batch of noise to train the generator based on the discriminator model every $N$ discriminator training rounds;
%\textcircled {6} Upload the local parameters to HAPS for model aggregation every $K$ local training rounds;
%\textcircled {7} HAPS obtains the global generator and discriminator models through the federated averaging;
%\textcircled {8} HAPS implements CA2C-based UAV selection to determine the optimal UAV sets for next round federated training;
%\textcircled {9} HAPS distribute model parameters and UAV selection indicators to all UAVs for their further updates.  The above steps are implemented repeatedly until the global models converge.

\section{AI for VHetNets: DRL-based intelligent and automatic network management}

Existing works mainly consider the anomaly detection as a standalone functionality that is independent of any other network management functionalities. However, the efficiency of the AFL-based anomaly detection scheme depends heavily on the network management functionalities of UAV selection, device association and UAV trajectory planing, whose functions are presented as follows:

\textbf{Device association.}  The device association strategy is applied to decide which UAV a ground IoT device should associate with to maximize its successful sensing probability and the whole networks' coverage, which are closely related to the data abundance for establishing an accurate and robust anomaly detection model.

\textbf{UAV selection.} The UAV selection strategy is used to select the UAV subset participating in each global aggregation round, whose objective is to minimize the federated execution time and learning accuracy loss. The federated execution time consists of local model update latency, local model upload latency, global model aggregation latency, and global model broadcast latency, which mainly rely on  UAVs' locations, remaining energy and computing resources.

\textbf{UAV trajectory planning.} Because a part of IoT devices randomly move with a rapid speed, UAVs also need to move correspondingly to complete sensing processes. However, the energy capacity of UAVs is limited, so the efficient UAV trajectory planning strategy is vital to balance the energy usage in flying for coverage and in computation and transmission for the training of the anomaly detection model.

Intelligent and automatic network management functionalities consist of sequential decision-making processes with the time-varying network environment. Reinforcement learning (RL) is an effective solution to make sequential decisions through unintermittent interactions with the network environment, which comprises a decision-making agent, network state, action, and reward \cite{24-9136602}.  Based on the  network state, the agent takes an action and then obtains an instant reward. The network environment will enter into a new state after an action is taken on it. The state transition and instant reward will guide and reinforce the adaption of the agent's policy, which determines the sequential actions taken in each decision period with the evolution of the network environment. The learning process continues until the optimal policy that maximizes the accumulated reward is found, which can be expressed by a state-value or an action-value function. The state-value function represents the total expected reward starting from an initial network state, while the action-value function, also known as the Q-value function, maps each state-action pair to the accumulated reward.

RL is hard to train to converge and obtain the optimal policy when facing large state and action spaces in large-scale networks. DRL overcomes this challenge by adopting deep neural networks (DNNs) as function approximators to predict different components in RL, including the value function, policy, and the network state transition model.

\subsection{DRL Approaches}
\textbf{DQN.} DQN is an extension of the traditional value-based Q-learning method, which utilizes DNNs as function approximators to predict Q-values. The success of DQN depends on two key techniques, namely fixed target Q-network and experience replay, to stabilize the learning process with large state and actions spaces. The fixed target Q-network means that parameters of the target network only update with Q-network parameters every $C$ steps and keep unchanged between two individual updates, so the challenge of non-stationary target values in DQN training can be overcome. The experience replay keeps a replay memory to buffer historical transition experience. In each DQN training round, a random mini-batch of samples are sampled from the replay buffer to break the sample correlation and improve the learning efficiency \cite{26-9154432}. The convergence of DQN training is achieved by minimizing the mean squared error between the Q-value estimated by Q network and the target value obtained from target Q-network. However, the DQN with fixed target Q-network and experience replay only can handle sequential decision-making processes with discrete action spaces.

\textbf{DDPG.} Policy gradient-based RL methods can be utilized to handle sequential decision-making processes with continuous action spaces by learning deterministic/stochastic policies. The classical actor-critic framework stems from the policy gradient principle by establishing connections between the Q-value and the policy gradient. The critic function evaluates the quality of the policy by estimating the Q-value and the actor function adopts the policy gradient method to update the policy parameters. To speed up the convergence of policy gradient-based methods, DDPG is proposed to combine the policy-based approach with the value-based approach to estimate the policy gradient more efficiently \cite{24-9136602}. To adapt to the complex networks with large state and action spaces, DNNs are used as function approximators for actor and critic networks. In addition, DDPG also uses two key techniques, namely target actor and critic networks, and experience replay to make the learning process more stable and efficient.

\textbf{CA2C.} As wireless networks become increasingly large-scale and complicated, the network management problems face kinds of decision variables, including both discrete indicators (such as device association, handover, sleep strategy, and so on) and continuous variables (such as power allocation, trajectory planning, and so on). Besides, the allocation of bandwidth, computing and storage resources can be considered as either discrete or continuous decision variables. Most of the time, discrete indicators and continuous variables are not independent and can be solved through establishing standalone decision-making models, but highly correlated to reach a common objective or multiple common objectives. Thus the CA2C approach is introduced to realize intelligent and automatic network management,which combines the advantages of DDPG approach for dealing with continuous decision variables and DQN approach for dealing with discrete decision variables \cite{26-9154432}.

\subsection{CA2C-based intelligent and automatic network management}

For the implementation of AI-native VHetNets-enabled anomaly detection framework, the decision-making processes of CA2C-based intelligent and automatic network management are defined as follows:

\textbf{Decision-making agent:} HAPS is worked as the decision-making agent due to its quasi-stationary position, the line-of-sight communication, wide coverage and multiple energy sources,  including conventional energy (such as electrical batteries and fuel tanks), energy beams, and solar energy \cite{21-9380673}.

\textbf{State:} In each decision period, HAPS interacts with the network environment to observe its network state, including current locations of ground IoT devices and UAVs, device association indicators and selected UAV subset in last decision period, and UAVs' remaining energy.

\textbf{Action:} HAPS is responsible for taking corresponding actions based on the observed network state, where actions include discrete actions of device association and UAV selection indicators, and continuous actions of UAV positions. The network environment will enter into a new state after corresponding actions are taken on the current network state.

\textbf{Reward:} The reward represents a numerical value obtained by the agent from the environment to quantify its satisfaction with taken actions. The objective of the defined decision-making processes is to maximize the coverage capacity for ground IoT devices while minimize the execution time and learning accuracy loss of the AFL-based anomaly detection model. Therefore, the instant reward is defined as the weighted sum of the coverage capacity, execution time and learning accuracy loss in one decision period.

Learning processes of the CA2C approach are illustrated in Fig. 3 and explained as follows, which include three main sub-procedures.

\begin{figure*}[htbp]
\centering
\includegraphics[width=5.5in]{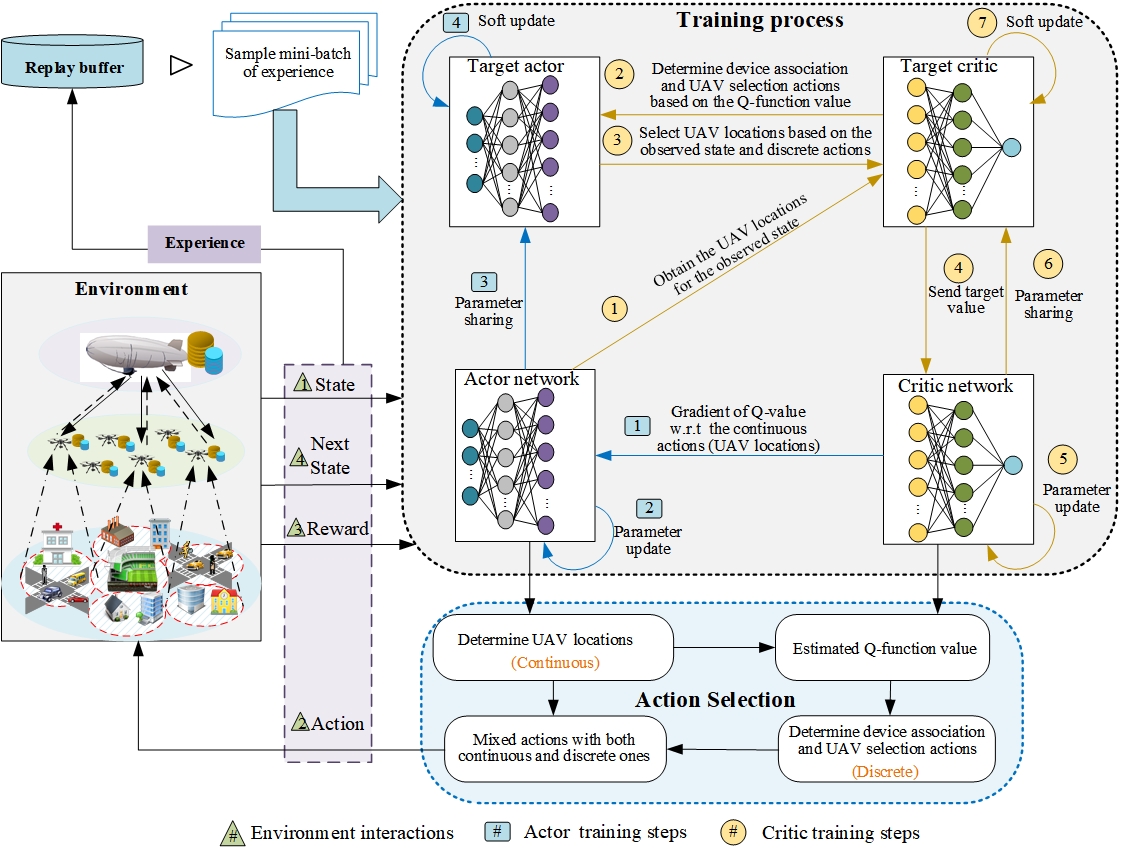}
\caption{Learning processes of the CA2C approach.}
\end{figure*}

\textbf{1. Environment interactions:}
\begin{enumerate}
    \item Observe the network state;
    \item  Execute mixed actions with both discrete and continuous ones determined by the actor and critic networks in each decision period;
    \item Calculate the instant reward as the feedback of selected actions;
    \item The environment enters into a new network state.
\end{enumerate}
The quadruple $<$State, Action, Reward, Next state$>$ will be stored in the replay buffer as experience.

\textbf{2. Critic network training:} First, sample a mini-batch of experience as the training data.
\begin{enumerate}
    \item Based on the actor network and the sampled experience, obtain the UAV locations (continuous actions) for the observed state;
    \item  Determine the device association and UAV selection actions (discrete actions) based on the Q-function value estimated by the target critic network;
    \item The target actor network calculates UAV locations based on the observed state and discrete actions;
    \item Calculate the target value by adding the instant reward in experience and the Q-function value estimated by the target critic network;
    \item Update parameters of the critic network using Adam optimizer;
    \item Share parameters of the critic network with the target critic network;
    \item Update parameters of the target critic network using the soft update method, namely the weighted sum of critic network's parameters and its own parameters, where the sum of weights is equal to 1.
\end{enumerate}

\textbf{3. Actor network training:} First, sample a mini-batch of experience as the training data.
\begin{enumerate}
    \item Calculate gradients of the Q-value function with respect to the UAV locations (continuous actions) for all sampled experience;
    \item  Update parameters of the actor network using Adam optimizer;
    \item Share parameters of the actor network with the target actor network;
    \item Update parameters of the target actor network using the soft update method, which is same as the update of the target critic network.
\end{enumerate}

The above three sub-procedures are executed alternately until all training episodes end. The integration between the AFL-based anomaly detection scheme and the CA2C-based intelligent and automatic network management approach contributes to an efficient AI-native  VHetNets-enabled anomaly detection framework.

\section{A case study}

In this section, a case study is presented to evaluate the performance of the proposed AI-native VHetNets-enabled anomaly detection framework.

We consider a 1 km $\times$ 1 km area with a HAPS, 5 UAVs and 30 ground IoT devices. The energy budgets of a UAV for each local model upload and each local training round are set as 30 J and 1 J, respectively. The flying energy consumption of a UAV is assumed to be proportional to the flying distance, whose rate is set as 300 J/km. The considered scenario is simulated in pytorch (Python 3.7) and the proposed scheme is conducted by a computer with a CPU capacity of 12 Intel(R) Core(TM) i7-10750H CPU $\times$ 2.6 GHz and a RAM of 16 GB.

We choose a well-known dataset published by Inter Berkeley Research Lab \cite{28584714} as sensing data from ground IoT devices. The dataset includes temperature, humidity, light and voltage features collected from 54 distributed Mica2Dot sensors every 31 seconds during the period of February 28th to April 5th, 2004. Each UAV trains its local anomaly detection model based on the sensing data from its associated IoT devices. HAPS is responsible for aggregating local models to obtain the global anomaly detection model every $K$ local training rounds, where $K$ is set as 30 in this article.

The AI-native VHetNets-enabled anomaly detection framework is established on the AFL-based GAN model, where the subset of UAVs that participate in every global training episode is determined by the CA2C approach. As shown in Fig. 4, we first present the convergence performance of global generator and discriminator in the AFL-based GAN model. In the simulation, socket is used to realize the communication between emulated UAVs and HAPS. It can be seen that the AFL-based GAN model has converged after 80 global training episodes.

\begin{figure}[!t]
\centering
\includegraphics[width=3.5in]{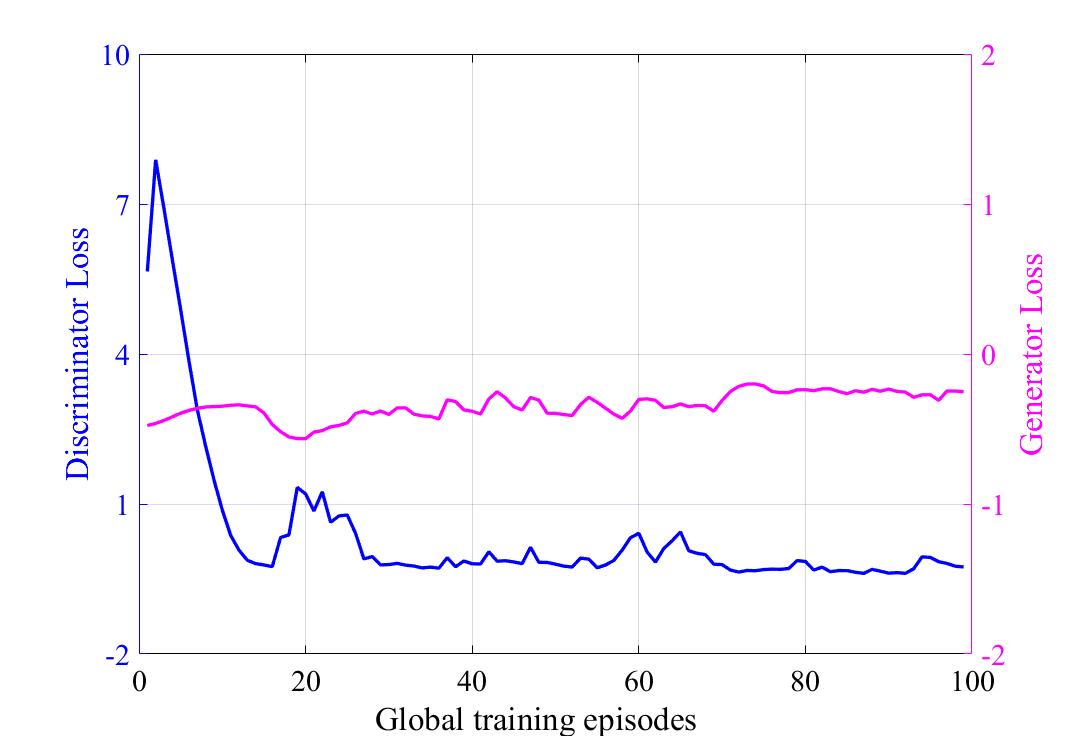}
\caption{Convergence performance of the AFL-based GAN model.}
\end{figure}

We have compared the detection performance and the energy consumption of the AFL-based GAN model with the FL-based GAN model and standalone GAN model. In FL-based GAN model, each UAV participates in all global aggregation rounds and the UAV selection process is not included. In standalone GAN approach, each UAV trains its own anomaly detection model without any data or parameter exchange. Metrics used for evaluating the detection performance include precision, recall, accuracy and F1-score. Their comparison results are presented in Fig. 5.

Fig. 5(a) shows the average energy consumption of UAVs as the global training proceeds. One global training episode includes 30 local training rounds. We can observe that the standalone GAN model consumes the minimum energy, and the AFL-based GAN model with UAV selection process can reduce the energy consumption compared to the FL-based GAN model without UAV selection process. The reason is that in the AFL-based GAN model, UAVs are not required to participate in every global aggregation, and hence more energy is saved. Fig. 5(b) shows the detection performance of different methods. It can be seen that the proposed model achieves the best detection performance compared to the FL-based GAN model and standalone GAN model. This is because the UAV selection process in the AFL-based GAN model can avoid UAVs with less learning accuracy participating in the global aggregation. Therefore, the AFL-based GAN model can obtain a more accurate anomaly detection model with less energy consumption in UAVs.

\begin{figure}[!t]
\centering
\subfigure[]{\includegraphics[width=3.3in]{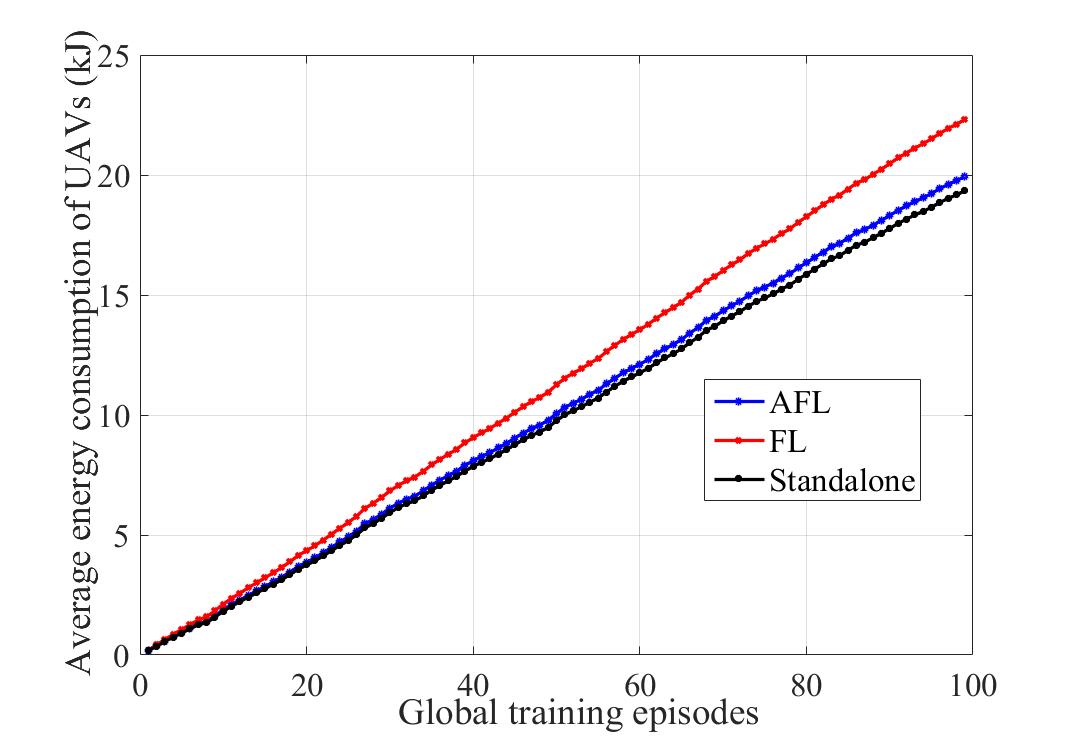}%
\label{fig_first_case}}
\hfil
\subfigure[]{\includegraphics[width=3.5in]{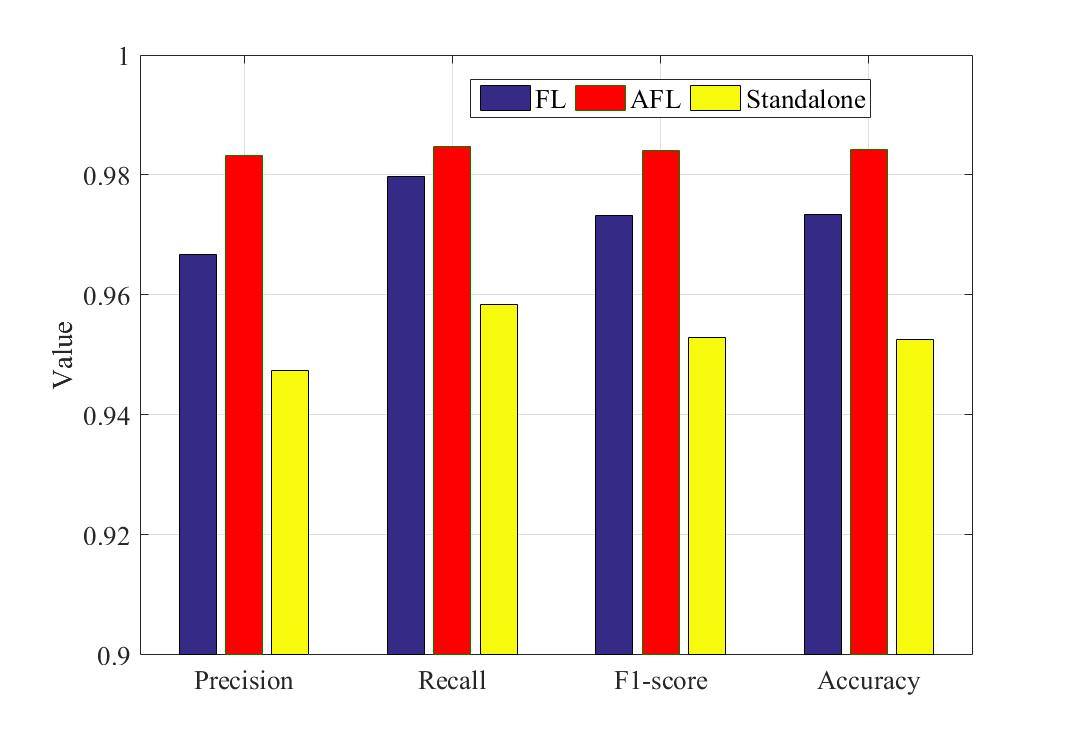}%
\label{fig_second_case}}
\caption{Performance comparison of different methods: (a) energy consumption; (b) detection performance.}
\label{fig_sim}
\end{figure}

\section{Conclusion and future research}
In this article, we have proposed an AI-native VHetNets-enabled anomaly detection framework for ubiquitous IoT through hierarchical coordination among HAPS, UAVs and ground IoT devices. The framework aims to enable the synergy of VHetNets and AI. VHetNets for AI considers VHetNets with communication and computation capacity as agents for distributed AI model training to provide the anomaly detection service. AI for VHetNets helps provide automatic and intelligent network management functionalities to facilitate the efficient implementation of anomaly detection service. The proposed AI-native VHetNets-enabled anomaly detection framework is established on the AFL-based GAN model, where the subset of UAVs that participate in every global training episode is determined by the CA2C approach. In the case study, we have validated that the proposed framework achieves a more accurate anomaly detection model with less energy consumption in UAVs.

For future research, the proposed framework can be extended by involving the blockchain technology to improve the privacy and security during distributed AI model training. First, although FL has distinct privacy advantages, it still faces two security threats: poisoning participants and inference attacks. These two threats may result in the convergence failure of global AI models or privacy leakage of sensitive information. Second, because UAVs and HAPS are operating in the sky and communicate through wireless communication technologies, they are more vulnerable to these privacy and security risks. Therefore, the recognition of malicious participants and secure transmission during aggregation are major concerns in constructing secure FL framework. With the advantages of decentralization, immutability, anonymity, and security, integrating blockchain with the FL framework can defend against numerous threats and attacks.

\section*{Acknowledgments}
This work was supported in part by the National Natural Science Foundation of
China under Grant Nos. 62071078 and 62001076, Sichuan Science and Technology Program under Grant No. 2021YFQ0053, and in part by Huawei Canada.

\bibliographystyle{IEEEtran}
 \bibliography{IEEEabrv,refrence}

\newpage

 %the LaTeX parser from getting confused when it sees the complicated
 %your own custom macro containing the $\backslash${\tt{includegraphics}} command to make things
 %simpler here.)


%\bf{If you include a photo:}\vspace{-33pt}
%\begin{IEEEbiography}[{\includegraphics[width=1in,height=1.25in,clip,keepaspectratio]{fig1}}]{Michael Shell}
%Use $\backslash${\tt{begin\{IEEEbiography\}}} and then for the 1st argument use $\backslash${\tt{includegraphics}} to declare and link the author photo.
%Use the author name as the 3rd argument followed by the biography text.
%\end{IEEEbiography}



\vfill

\end{document}